\documentstyle[aps,prl,twocolumn,epsf]{revtex}
\input psfig
\begin{document}
\title{Nonuniversality of weak synchronization in chaotic systems}
\author{Maria de Sousa Vieira\cite{email} and Allan J. Lichtenberg} 
\address{Department of Electrical Engineering and Computer Sciences, University of California, Berkeley, CA 94720, USA.}
\maketitle
\begin{abstract}
We show that the separate properties of weak synchronization (WS)   
and strong synchronization (SS), reported recently 
by Pyragas [K. Pyragas, Phys. Rev. E, {\bf 54}, R4508 (1996)],  in 
unidirectionally coupled chaotic systems,  are 
not generally distinct properties of such systems. In particular, 
we find analytically for the tent map and numerically for some 
parameters of the circle map that the transition to WS and SS
coincide. 
\end{abstract} 
\narrowtext
Chaotic systems, by definition, are characterized by strong sensitivity 
to the initial conditions. Thus, in a general situation, one cannot  
synchronize two chaotic systems, since in a practical situation it is 
impossible to start the evolution of the two systems with {\sl exactly}
the same initial conditions. However, one can synchronize subsystems of 
a chaotic system that have a chaotic output, but are intrinsically stable. 
This was shown  by  
Pecora and Carroll\cite{pecora}.   
More specifically, Pecora and Carroll studied  dynamical systems of 
the   
type $\dot u=g(u,w)$, $\dot w=h(u,w)$ and showed that     
variable $\dot w'$ governed by $\dot w'=h(u,w')$ can   
synchronize with $w$ if the sub-Liapunov exponents of the driven 
subsystem $w'$ are all negative (implying that the subsystem $w'$ is
stable). The sub-Liapunov exponents 
they defined depend on the Jacobian matrix of the 
$w$ subsystem, taking derivatives with respect to $w$ only. 
 
In a recent letter, Pyragas\cite{pyragas} studied one-dimensional 
chaotic systems, governed by a function $f$, 
coupled unidirectionally in the following way  
\begin{eqnarray}
x_{n+1}&=&f(x_n) \nonumber \\
y_{n+1}&=&f(y_{n}) - k[f(y_{n})-f(x_{n})]  
\label{eq1} \\ 
z_{n+1}&=&f(z_{n}) - k[f(z_{n})-f(x_{n})]. \nonumber
\end{eqnarray}
He found numerically that in a specific system, namely, 
the logistic map, synchronization 
between the variables $x$, $y$ and $z$, as $k$ is increased,  
occurs in two stages. 
In the first stage, $y$ synchronizes with $z$ and not with $x$, starting 
at a critical 
value $k_w$. He called this {\sl weak synchronization} (WS). 
The second stage of synchronization starts at $k=k_s$, with $k_s > k_w$, and 
is characterized by the synchronization of $x$, $y$ and $z$,  
which he called {\sl strong synchronization} (SS). 
Note that in above equation the case $k=1$ represents synchronization 
in a trivial way, since this leads to $y_{n+1}=z_{n+1}=x_{n+1}$. 
So, for this kind of coupling $k$ is taken in the interval $[0,1)$.  

The conditions under which  WS and SS occur  
were determined by two 
Liapunov exponents, that is, the conditional Liapunov exponent, 
\begin{equation}
\lambda^R=\ln(1-k)+\lim_{N\to\infty} {{1}\over{N}} \sum_{n=1}^N \ln|f'(y_n)|, 
 \label{eq2}
\end{equation}
defining the stability of the invariant manifold $y=z$, and the transverse 
Liapunov exponent of the invariant manifold $x=y$, 
\begin{equation}
\lambda_0=\ln(1-k)+\lim_{N\to\infty} {{1}\over{N}} \sum_{n=1}^N \ln|f'(x_n)|. 
 \label{eq3}
\end{equation}
In  the logistic map, Pyragas found that 
$\lambda^R $ becomes zero at two characteristic values of the coupling 
strength, $k_w$ and $k_s$, corresponding to the threshold of WS and SS, 
respectively. In the region of weak synchronization ($k_w < k < k_s$), 
$\lambda^R < 0$ and 
$\lambda_0 > 0$.   
Strong synchronization occurs for $k>k_s$, where these two 
Liapunov 
exponents 
coincide, i.e., $\lambda _0=\lambda ^R$. 

We studied the phenomenon of chaotic synchronization in other one-dimensional 
maps, coupled also according to Eq.~(\ref{eq1}). We found that 
the phenomenon of weak synchronization is not always found in 
such systems.  

We start by showing what is the relationship between the exponents 
$\lambda^R$ and $\lambda_0$, defined by Pyragas, and the Liapunov  
exponents of the three dimensional system (with dynamical variables 
$x$, $y$ and $z$) governed by Eq.~(\ref{eq1}). We will call this the global 
system, which     
has the following Jacobian matrix at a single position along the orbit:
\begin{equation}
 J_n =  \left( \begin{array}{cccc}
                f'(x_n) & 0 & 0 \\
                kf'(x_n) & [1-k]f'(y_n) & 0 \\
                kf'(x_n) & 0 & [1-k]f'(z_n) 
                \end{array}
         \right).  
    \label{eq4}
\end{equation}
The Liapunov exponent of this system is found by calculating the 
eigenvalues of the matrix that consists of the product of the 
Jacobian matrices along a given orbit. 
It turns out that, because of the symmetry of this particular 
matrix, the product $J$ of the Jacobian matrices $J_n$ is
of the type 
\begin{equation}
 J =  \left( \begin{array}{cccc}
                a_{11} & 0 & 0 \\
                a_{21} & a_{22} & 0 \\
                a_{31} & 0 & a_{33} 
                \end{array}
         \right), 
     \label{eq5}
\end{equation} 
where $a_{11}$, $a_{21}$, $a_{22}$, $a_{31}$ and $a_{33}$ are in principle 
different from zero. 
The eigenvalues of this matrix are 
\begin{equation}
\Lambda_1=a_{11}=\prod _{n=1}^N f'(x_n),  
 \label{eq6}
\end{equation}
\begin{equation}
\Lambda_2=a_{22}=\prod _{n=1}^N [1-k]f'(y_n),  
 \label{eq7}
\end{equation}
\begin{equation}
\Lambda_3=a_{33}=\prod _{n=1}^N [1-k]f'(z_n).   
 \label{eq8}
\end{equation}
Consequently, in 
this way of coupling, the eigenvalues  of the system are easily 
found, and each one is a function of a single variable.   
The Liapunov exponents are therefore  
\begin{equation}
\lambda_1=\lim_{N\to\infty} {{1}\over{N}} \sum_{n=1}^N \ln|f'(x_n)|, 
 \label{eq9}
\end{equation}
\begin{equation}
\lambda_2=\ln(1-k)+\lim_{N\to\infty} {{1}\over{N}} \sum_{n=1}^N \ln|f'(y_n)|, 
 \label{eq10}
\end{equation}
\begin{equation}
\lambda_3=\ln(1-k)+\lim_{N\to\infty} {{1}\over{N}} \sum_{n=1}^N \ln|f'(z_n)|. 
 \label{eq11}
\end{equation}
If the initial values of $y$ and $z$ are in the same basin of 
attraction,   
then $\lambda_3=\lambda_2$, because the parameters of the maps are the 
same. 
The Liapunov exponent $\lambda ^R$   
is equal to one of the Liapunov exponents 
of the global system, namely, it is equal to 
$\lambda_2$. 
Thus, this Liapunov exponent has a clear physical importance 
even when $y$ and $z$ are not synchronized. 
On the other hand,  
$\lambda_0$ is not a Liapunov exponent of the global system in the 
region where $x$ and $y$ are not synchronized (weak synchronization).
However, $\lambda_0$ and $\lambda_1$ are related to each other 
via a trivial additive term, namely, $\lambda_0=\ln(1-k)+\lambda_1$.   
It is obvious that in the region of SS, $\lambda_0=\lambda^R=\lambda_2$.  

We found in \cite{maria} that the loss of synchronization between 
subsystems of a dynamical system coupled according to Pecora and Carroll's  
coupling is characterized by the  
the Liapunov exponents of the global system. That is, when  
one of such Liapunov exponents crosses zero 
(in the positive direction) the two 
identical subsystems  
will lose synchrony. When this occurs, a transition from chaos to 
hyperchaos\cite{rossler}  takes place.  
Here we find a  similar phenomenon.  
The transition from SS to WS corresponds to the transition from chaos 
to hyperchaos in the global system.  

Our next step is to show that WS does not necessarily precede SS. We show 
this analytically  in the tent map, and later numerically in the circle map.

The tent map is defined as 

\begin{equation}
f(x)= \left\{ \begin{array} {ccc}   
                         ax,& {\rm if}\ 0 \le x \le 1/2,\\
                         a(1-x) ,& {\rm if} \ 1/2 \le x \le 1.
               \end{array}
  \right. 
 \label{eq12}
\end{equation}
The 
tent map has a period-one orbit when 
$0\le a < 1$, with fixed point 
$x^*=0$. For $1 < a < 2$ the map has a chaotic orbit. For $a>2$ the 
orbit diverges. (Note 
that, to avoid divergencies when $0 \le a < 2$, 
the initial conditions must be 
in the interval $[0, 1]$). Since $|f'(x_n)|=a$ in any point of 
the orbit, we have, using Eqs.~(\ref{eq9})-(\ref{eq11})   
\begin{equation}
\lambda_1= \ln a, 
 \label{eq13}
\end{equation}
\begin{equation}
\lambda_2=\lambda _3=\ln(1-k)+ \ln a. 
 \label{eq14}
\end{equation}
For a fixed $a$, we find that 
$\lambda_2$ decreases monotonically 
as $k$ is increased from $k=0$. Consequently, there is no 
region of WS in this map. For this map, 
synchronization between $x$ and $y$ and 
$z$ occurs simultaneously, and only what is called SS is seen. The transition 
where SS occurs is determined by $\lambda_2 <0$, which 
gives  $k_s = 1-1/a$.

Next, we consider coupled circle maps according to Eq.~(\ref{eq1}), 
where   
\begin{equation}
f(x)=x+\omega-{{b}\over{2\pi}}\sin(2\pi x).
 \label{eq15}
\end{equation}
Here we find regions of the parameter space where  WS is 
not observed. 
We show in Fig.~\ref{f1}(a) an example for this, where 
$b=6$ and $\omega=0.44$. 
The solid line in that figure 
represents $\lambda ^R$, which is equal to $\lambda _2$, and 
the dashed line represents $\lambda _0$, which is equal to   
$\log (1-k) + \lambda _1$. We do not find a region of $k$ 
where $\lambda ^R <0 $ and $\lambda _0 >0 $, and consequently 
no WS is seen. 

For the circle map (and also in the logistic map) we 
observed  interesting phenomena when the driving variable $x$   
is in   
one of the periodic windows of the chaotic band.  
There, we see regions 
of $k$ where $y$ and $z$ are chaotic ($\lambda_2 = \lambda ^R >0$) even 
when $x$ is periodic ($\lambda _1 <0$  and $\lambda _0 <0$). 
This is shown in Fig.~\ref{f1}(b), where $b=4$ and $\omega =0.4$. 

In \cite{pyragas}, Pyragas also studied the coupling of the 
Rossler's and Lorenz's systems according to 
\begin{eqnarray}
\dot x_1&=&-\alpha [x_2+x_3], \nonumber \\
\dot x_2&=&\alpha [x_1+0.2x_2],  
\label{eq16} \\
\dot x_3&=&\alpha [0.2+x_3(x_1-5.7)], \nonumber
\end{eqnarray} 
\begin{eqnarray} 
\dot y_1&=&10(-y_1+y_2), \nonumber \\
\dot y_2&=&28y_1-y_2-y_1y_3+kx_2,  
\label{eq17}\\  
\dot y_3&=&y_1y_2-8/3y_3, \nonumber 
\end{eqnarray}  
where $\alpha=6$. 
Here, we also can understand the phenomenon of 
weak synchronization reported by Pyragas for 
this coupled system, by considering it as a single global system of six 
variables.  We calculated all the Liapunov exponents of the 
global system, and found that at $k\approx 6.6$ one of the Liapunov 
exponents change sign. Beyond this critical value of 
$k$, synchronization between two copies of 
the Lorenz system [Eq.~(\ref{eq17})] with different initial conditions 
(in the same basin of attraction) will occur. 
That is, WS will be seen. No other change of 
sign was found in the Liapunov exponents as we increased $k$  
to $k=200$. 
Next, we calculated the information dimension $D_i$ of the attractor 
of the global system, assuming that the Kaplan and Yorke\cite{kaplan}
conjecture 
holds. Contrarily to the numerical results reported in 
\cite{pyragas}, we 
find numerically that the dimension of the global attractor does not 
converge 
to the dimension of the driving system [Eq.~(\ref{eq16})]. 
The convergence of these two dimensions was 
considered in \cite{pyragas} as the characterization of SS in 
this system. 
As Fig.~\ref{f2} shows, for $k \gtrsim 50$, $D_i$ 
for the global attractor (solid line) 
is approximately constant, but different from the information 
dimension of the driving system (dashed line).
For the global system we find that $D_i = 2.16$ when 
$k \gtrsim 50$ and for the driving system $D_i = 2.01$ (with the last 
digit being uncertain in both cases). 
Consequently, although we observed the property of WS 
in the coupled system governed by Eqs.~(\ref{eq16}) and (\ref{eq17}), 
we have not identified a regime of SS, as characterized in 
\cite{pyragas}. 

In summary, we have found that the property of weak synchronization 
in maps coupled according to Eq.~(\ref{eq1}) is a only a particular 
property 
of some systems. We have found that it does not hold in at least 
two systems, namely, the circle map and the tent map. We have 
also clarified the relationship between  
$\lambda ^R$ and $\lambda _0$ and the Liapunov exponents of the 
global system. Finally, we have verified that strong synchronization 
as defined in \cite{pyragas} is not observed in the coupled system 
governed by Eqs.~(\ref{eq16}) and (\ref{eq17}).

\begin{figure}
\caption[f1]{
Liapunov exponents (in Pyragas' notation) $\lambda ^R$ (solid line) and 
$\lambda _0$ (dashed line) for coupled circle maps, with (a) $b=6$ and 
$\omega=0.44$, and 
(b) $b=4$ and $\omega =0.4$. We used $N=30,000$ in 
Eqs.~(\ref{eq2}), (\ref{eq3}), and neglected a transient of 3,000 iterations. 
The initial conditions were $x=0.1$, $y=0.2$ and $z=0.3$. 
}
\label{f1}
\end{figure}  

\begin{figure}
\caption[f2]{Information dimension $D_i$ as a function of $k$ for the 
global attractor [Eqs.~(\ref{eq16}) and (\ref{eq17})] (solid line) and 
for the driving system [Eqs.~(\ref{eq16})] (dashed line). The error 
bars for $D_i$ are smaller than the diamond symbol. Since as 
$k$ increases, the variables $y_i$ change faster than $x_i$ we  
decrease the integration time step $\Delta t$ according 
to $\Delta t=0.02/(k+1)$. The initial conditions used were $x_1=0.1$, 
$x_2=0.2$, $x_3=0.3$, $y_1=0.4$, $y_2=0.5$ and $y_3=0.6$.      
}
\label{f2}
\end{figure}

\newpage
\begin{figure}
\centerline{\psfig{figure=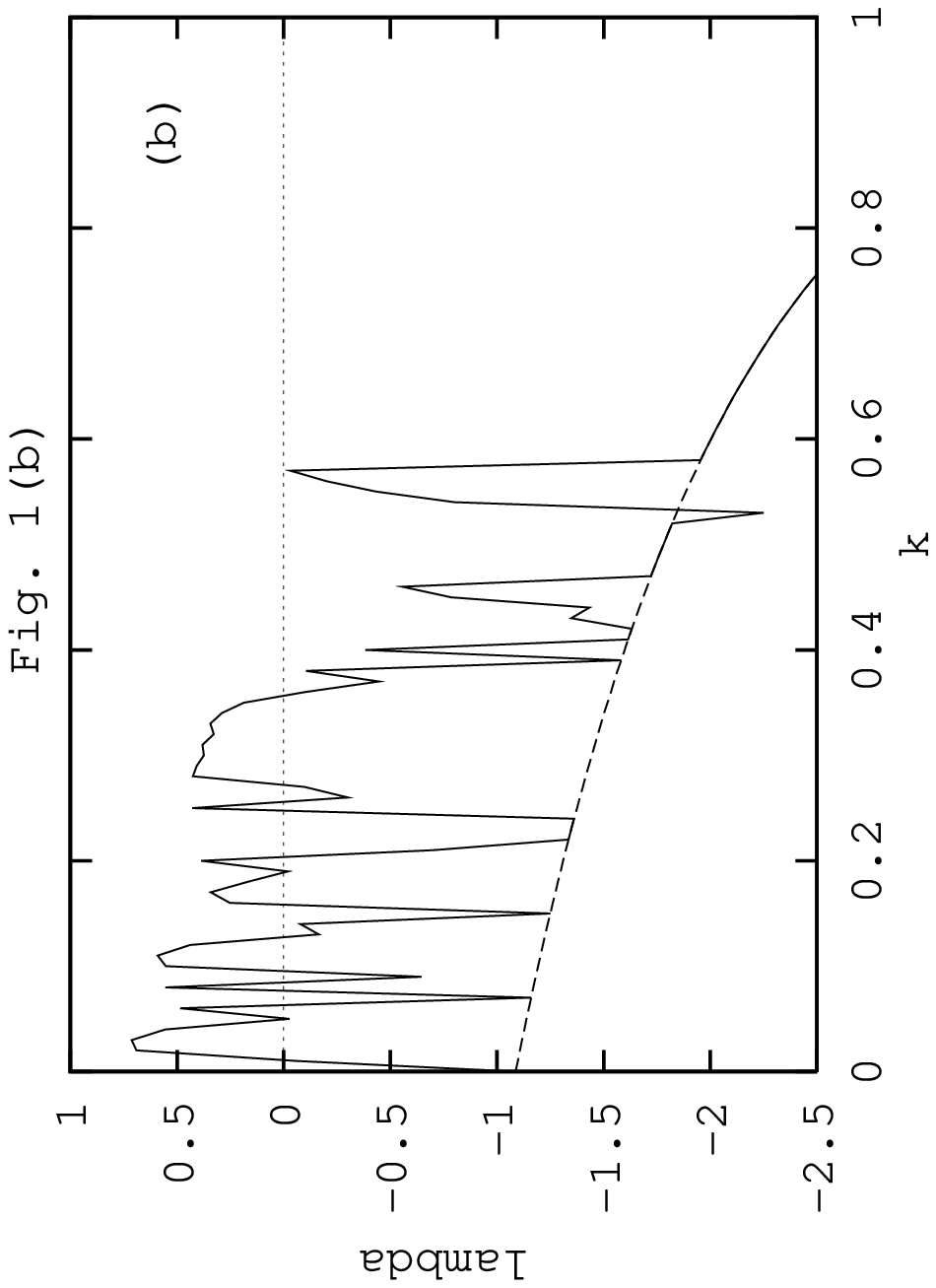,height=8.0cm}}
\end{figure}
\vskip 2cm
\begin{figure}
\centerline{\psfig{figure=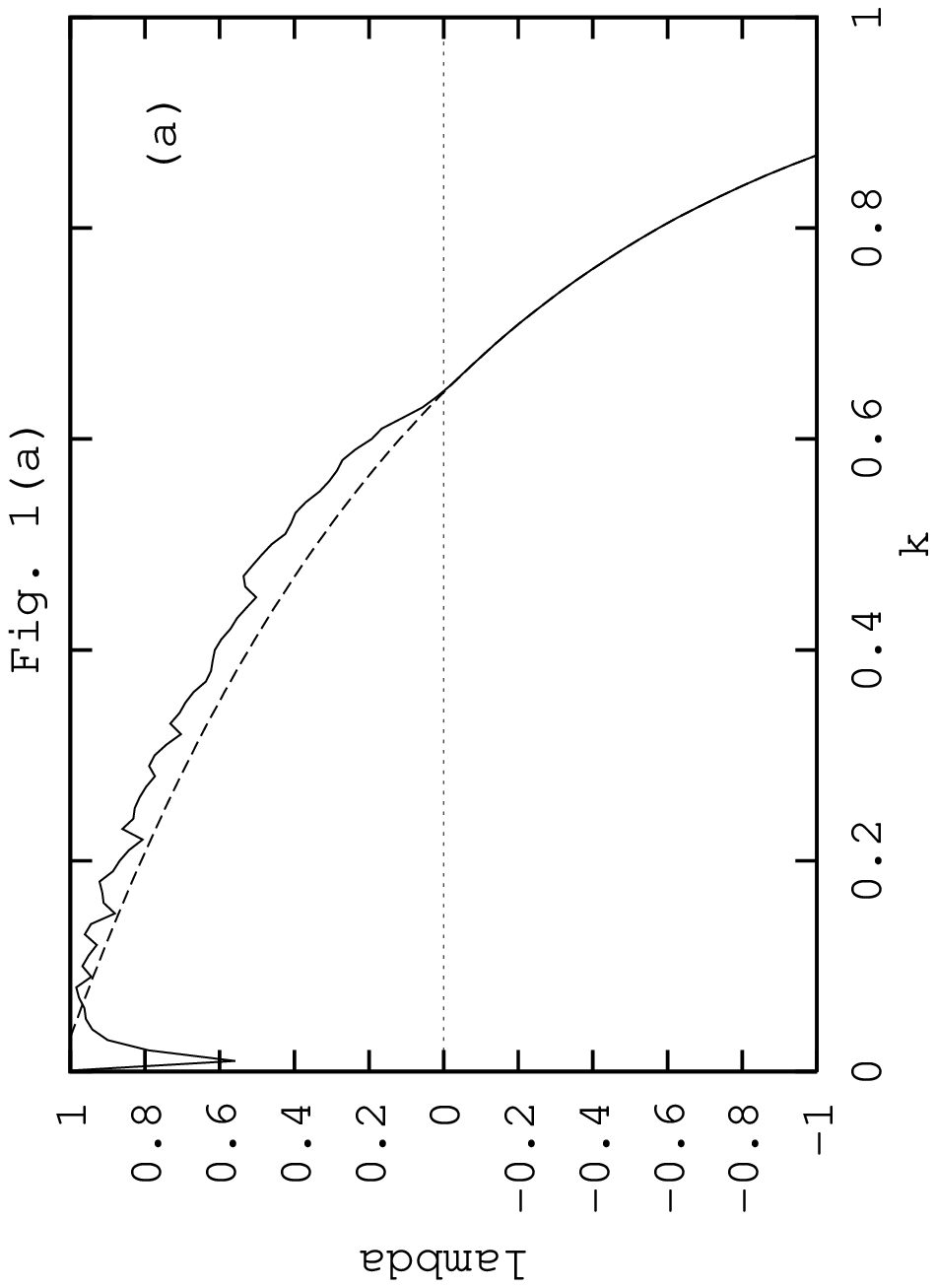,height=8.0cm}}
\end{figure}
\newpage
\begin{figure}
\centerline{\psfig{figure=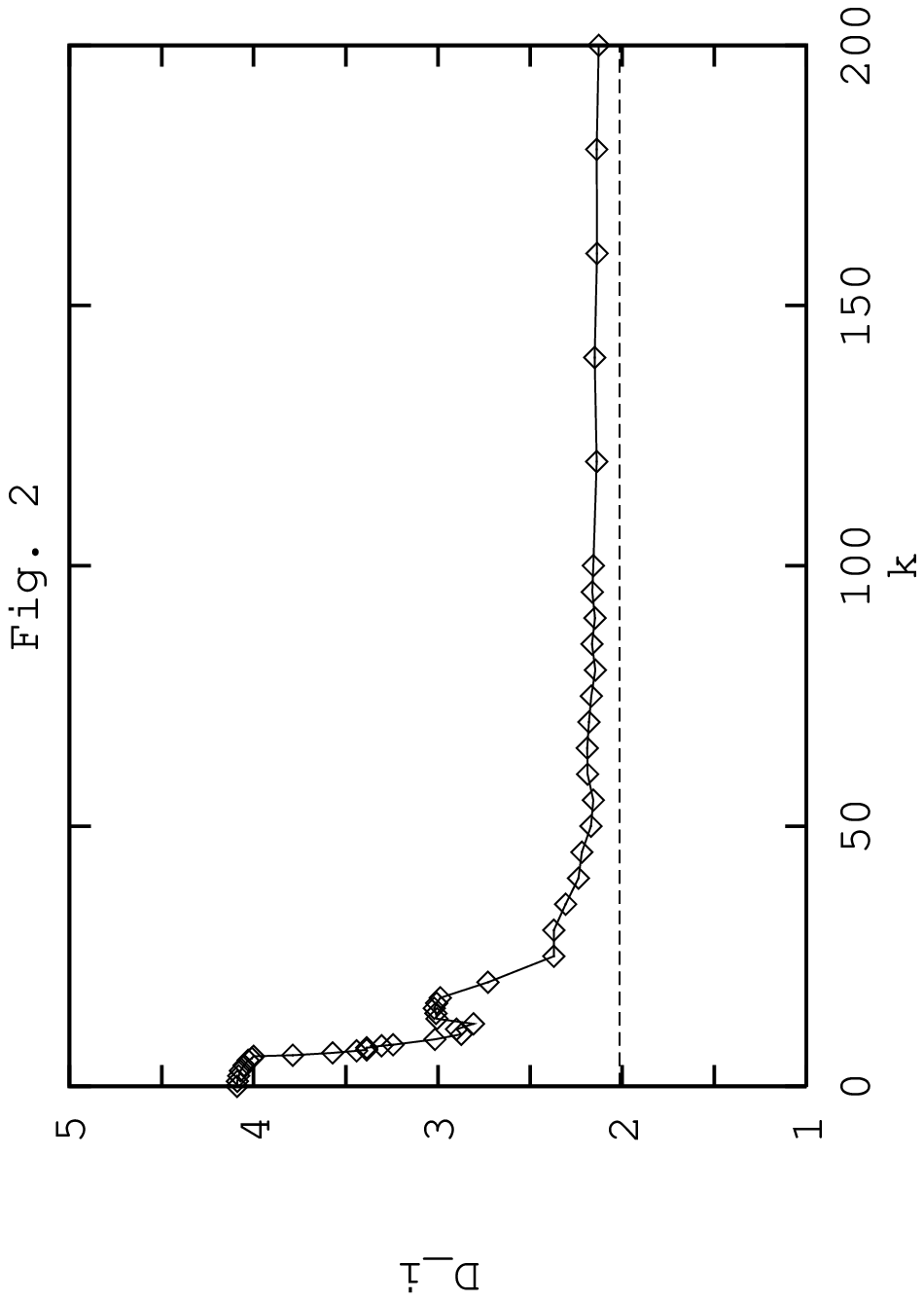,height=8.0cm}}
\end{figure}
\end{document}